# Electrically driven dynamic three-dimensional solitons in nematic liquid crystals

Bing-Xiang Li[1], Volodymyr Borshch[1], Rui-Lin Xiao[1], Sathyanarayana Paladugu[1], Taras Turiv[1], Sergij V. Shiyanovskii[1], and Oleg D. Lavrentovich[1,2,*]

[1]Liquid Crystal Institute and Chemical Physics Interdisciplinary Program, Kent State University, Kent, OH, 44242, USA

[2]Department of Physics, Kent State University, Kent, OH, 44242, USA

*olavrent@kent.edu

**Abstract**:

Electric field induced collective reorientation of nematic molecules placed between two flat parallel electrodes is of importance for both fundamental science and practical applications. This reorientation is either homogeneous over the area of electrodes, as in liquid crystal displays [1], or periodically modulated, as in the phenomenon called electroconvection[1,2] similar to Rayleigh-Bénard thermal convection[3]. The question is whether the electric field can produce spatially localized propagating solitons of molecular orientation. Here we demonstrate electrically driven three-dimensional particle-like solitons representing self-trapped waves of oscillating molecular orientation. The solitons propagate with a very high speed perpendicularly to both the electric field and the initial alignment direction. The propulsion is enabled by rapid collective reorientations of the molecules with the frequency of the applied electric field and by lack of fore-aft symmetry. The solitons preserve spatially-confined shapes while moving over distances hundreds of times larger than their size and survive collisions. During collisions, the solitons show repulsions and attractions, depending on the impact parameter. The solitons are topologically equivalent to the uniform state and have no static analogs, thus exhibiting a particle-wave duality. We anticipate the observations to be a starting point for a broad range of studies since the system allows for a precise control over a broad range of parameters that determine the shape, propagation speed, and interactions of the solitons.

==========================

Solitons are localized waves which propagate at a constant speed without changing their shape thanks to a balance of dispersive and nonlinear effects. Solitons preserve their identities after pairwise collisions, thus behaving as particles. Some media, such as nematic liquid crystals, offer especially rich opportunities for their studies. A uniaxial nematic is a medium with a long-range orientational order of molecules specified by a director $\hat{\mathbf{n}}$, with the properties $\hat{\mathbf{n}} \equiv -\hat{\mathbf{n}}$, $|\hat{\mathbf{n}}|^2 = 1$. The orientational order makes the nematic fluid anisotropic with a strong coupling between $\hat{\mathbf{n}}$ and velocity and capable of a nonlinear response to external fields. Exploration of director perturbations as solitons started about 50 years ago [4], facilitated by the fact that the director is the optic axis thus its orientation can be characterized in details by optical microscopy. Early studies focused on one-dimensional (1D) solitons representing domain walls[4], while the current research deals with 2D optical solitons-"nematicons"[5]. Nematicons are self-focused laser beams propagating in a nematic. They are stable against perturbations in the transverse direction but are not self-trapped along the direction of propagation. Multidimensional 3D particle-like solitons in nematics are deemed unstable with respect to shrinking, since a decrease in size, $L \to \mu L$ by a factor $\mu < 1$ entails a decrease in the total elastic energy, $F \to \mu F$ [6]. Static 3D solitons can be stabilized in cholesterics, which are chiral nematics with helical twisting of the director[7-10]. Their stability is topologically protected by the fixed pitch of the helicoid[7]. These solitons do not require motion for self-entrapment. In other media, the solitons are mostly 1D (such as surface waves in shallow channels) or 2D (optical solitons-beams[11]); recently, trains of 3D optical solitons have been demonstrated[12].

We present an experimental observation of 3D solitons with dual particle-wave character that propagate through a nematic powered by an alternating current (AC) electric field. They are topologically equivalent to a uniform state and have no static analogs. The solitons maintain their compact shape along all three coordinates. Within the soliton, the director perturbations oscillate with the AC electric field and break the fore-aft symmetry of the structure, which results in rapid propagation. The solitons survive collisions and show short-range interactions.

We used a nematic 4'-butyl-4-heptyl-bicyclohexyl-4-carbonitrile (CCN-47) with a negative anisotropy of permittivity, $\Delta\varepsilon = \varepsilon_\| - \varepsilon_\perp < 0$ and conductivity, $\Delta\sigma = \sigma_\| - \sigma_\perp < 0$ [13]; the subscripts indicate whether the component is measured along $\hat{\mathbf{n}}$ or perpendicularly to it. The nematic is aligned uniformly, $\hat{\mathbf{n}}_0 = (1,0,0)$, in flat cells of thickness $d = (3-30)\,\mu\text{m}$, Fig.1a. A sinusoidal field $\mathbf{E} = (0,0,E)$ of frequency $f = 20\,\text{Hz} - 5\,\text{kHz}$ is perpendicular to the $xy$ plane of the cell.

**Nucleation and structure of solitons.** As the voltage increases above some frequency-dependent threshold $U_{soliton}$, 3D localized director distortions nucleate and move perpendicularly to $\hat{\mathbf{n}}_0$ and $\mathbf{E}$, over distances much longer than their size, without spreading and surviving collisions, thus representing solitons, Fig. 1b-e. Viewed under a polarizing microscope with one of the polarizers along $\hat{\mathbf{n}}_0$, a soliton resembles a "flying tuxedo", Fig.1b-e. Outside the tuxedo, $\hat{\mathbf{n}}$ remains parallel to the $x$-axis, and the intensity of transmitted light is close to 0. Inside the tuxedo, transmission increases, indicating azimuthal deviations of $\hat{\mathbf{n}}$ by some angle $\varphi \approx 20° - 35°$, Fig.1b-e. When a red plate compensator (530 nm) is inserted with the optic axis $\lambda$ making 45° with the polarizer, one shoulder of the soliton appears yellow and the other blue, Fig.1c-e. In the blue region, $\hat{\mathbf{n}}$ tilts toward $\lambda$, in the yellow region, it tilts away from $\lambda$ [6], compare Fig.1b to Fig.1c,d,e. The in-plane director within the tuxedo resembles a bow, Fig.1b.

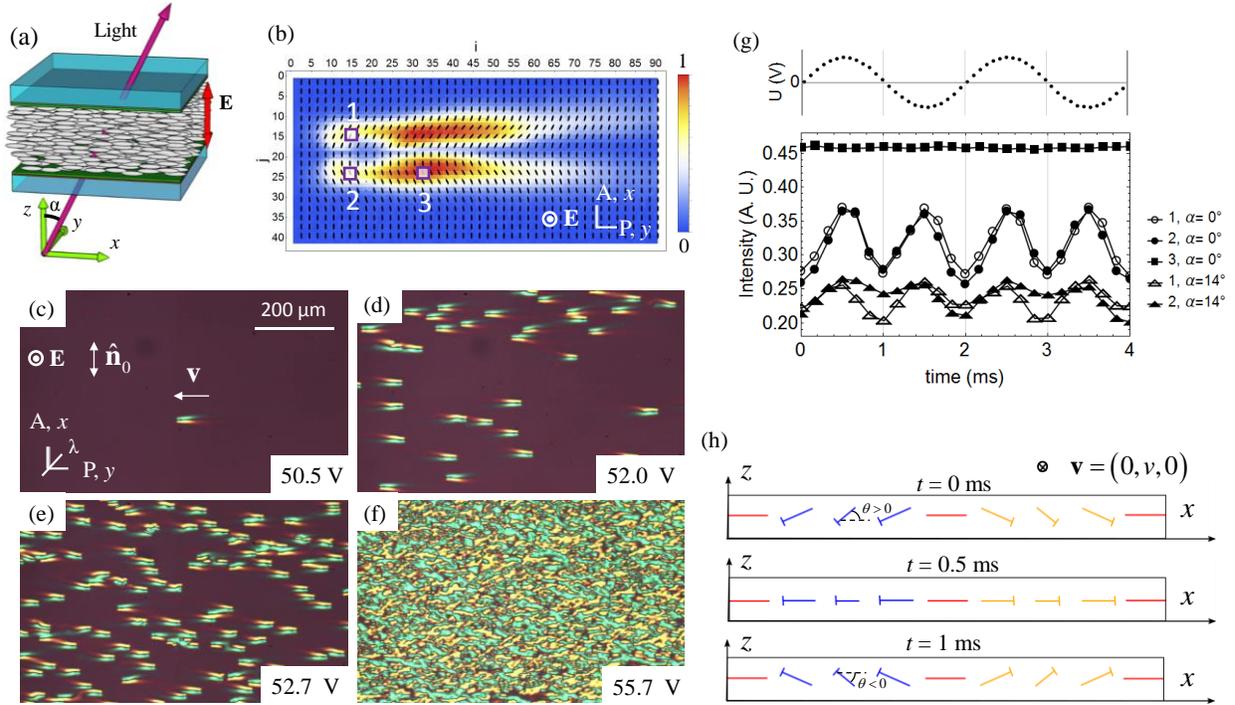

**Fig. 1 Dynamic solitons in a planar nematic cell.** (a) Cell scheme. (b) Transmitted light intensity map and director distortions in the $xy$ plane within a soliton ($U = 65.4$ V, $f = 500$ Hz, $T = 40\,°C$, $d = 8.0\,\mu m$). The bar shows a linear scale of transmitted light intensity; the pixels are labelled by (i,j) coordinates. (c,d,e) Solitons observed between crossed polarizers and a compensator with the optic axis $\lambda$ as the voltage $U$ increases as indicated ($f = 450$ Hz, $T = 45\,°C$, $d = 7.5\,\mu m$). (f) Stripe domains formed at an elevated voltage $U = 55.7$ V. (g) Dynamics of transmitted light intensity for the same parameters as in part (b), at locations 1, 2, and 3 indicated in panel (b) for normal, $\alpha = 0$, and oblique, $\alpha = 14°$, incidence of light. (h) Schematic 3D director field at the head of the soliton for three instants separated by a quarter-period of the applied voltage. The nails represent the tilted director with the heads closer to the observer than the ends.

At a fixed $d$, the solitons preserve their width $w \approx 2d$ along the $x$-axis; $w$ does not change with $f$ and the voltage $U$, Fig.1c-e. The length $L$ of the solitons, defined as the distance between the point of maximum light intensity and a point towards the tail where the transmitted intensity is smaller by a factor $e \approx 2.7$ is in a range $L \approx (20-50)$ μm and increases with $U$ (Extended data Fig.1). Both $w$ and $L$ are orders of magnitude smaller than the system lateral extensions of 5 mm.

The director in the left and right sides of a soliton synchronously tilts up and down, by an angle $\theta$ (measured between $\hat{\mathbf{n}}$ and the $xy$ plane), thus causing periodic modulations of light transmission, Fig.1g. The oscillations of $\theta$ are localized at the head of the tuxedo, coordinates i=7-25 in Fig. 1b, with the maximum $|\theta| \approx 20° - 35°$ in the pixels labelled 1 and 2 in Fig.1b. The polar tilts are maximum in the middle plane of the cell (Extended Data Fig.2). In the rest of tuxedo, i > 25 and pixel 3 in Fig.1b, the light intensity does not oscillate, thus $\theta = 0$, Fig.1g.

To characterize better the period and polarity of oscillations, we used oblique incidence, with an angle $\alpha \approx 14°$ between the incident light and the $z$-axis, Fig.1a. In this case, the up and down tilts of $\hat{\mathbf{n}}$ produce different projections onto the polarization direction of light (the $y$-axis), Extended data Fig.3. The light intensity variations for pixels 1 and 2 in Fig.1g show that $\theta$ oscillates with the same frequency $f$ as the applied field, Fig.1h. There is a phase shift between $E$ and $\theta$. The oscillations of $\theta$ do not change substantially the azimuthal angle $\varphi$ in the coordinate frame moving with the soliton.

The solitons nucleate at irregularities such as surface imperfections, dust particles and edges of the electrodes (Extended data Fig.4,5). They exist in a broad range of thickness, $d = (3.0-19.5)$ μm, frequency, $f \approx (0.1-4)$ kHz, and temperature, $T = (30-58)$ °C. The frequency range is limited from below by $f_{soliton}$ that decreases as $d$ increases. At 45 °C, $f_{soliton}$ = 550 Hz and 130 Hz, in cells with $d = 3.5$ μm and 9.3 μm, respectively. At a fixed $f$, the solitons exist in a narrow range of voltages $U = (1-1.1)U_{soliton}$, Fig.2a. The number of solitons increases with $U$, Fig.1c,d,e. Once the voltage exceeds another threshold $U_{stripes} \approx 1.1 U_{soliton}$, the solitons are replaced by an array of periodic stripes, Fig.1f. Both $U_{soliton}$ and $U_{stripes}$ increase with $f$, Fig.2a. In thick cells, $d > 19.5$ μm and low frequencies, $f < f_{soliton}$, the uniform state changes directly to the periodic stripes.

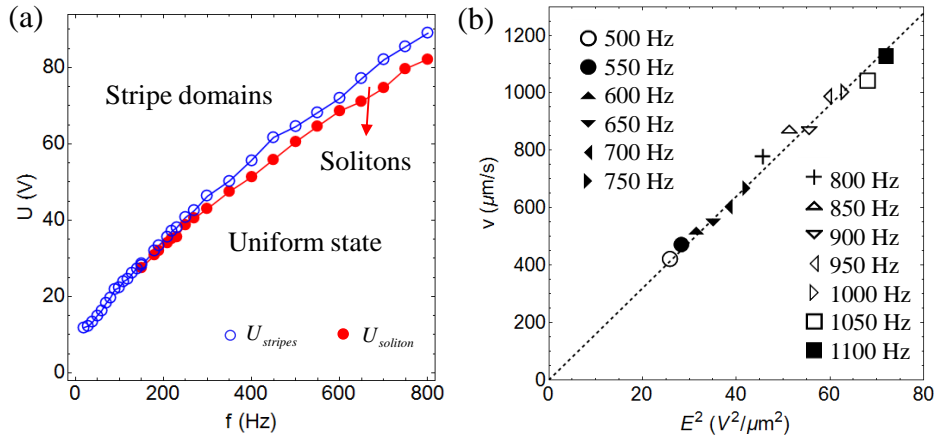

**Fig. 2 Soliton properties as a function of the applied field.** (a) Frequency dependencies of $U_{soliton}$ and $U_{stripes}$ (8.0 μm, 45 °C). (b) Velocity of solitons vs. square of the electric field (7.8 μm, 50 °C); the applied field corresponds to the frequencies specified in the legend.

The velocity $v$ of the solitons is huge, $v \approx (10-50) L/s$, and grows as $v = \beta E^2$, where $\beta \approx 1.5 \times 10^{-17}$ V$^{-2}$m$^3$s$^{-1}$, Fig.2b. This dependence is indirect, since for a fixed $f$, the solitons exist only within a narrow range of voltages. Fig 2b shows the data for different frequencies.

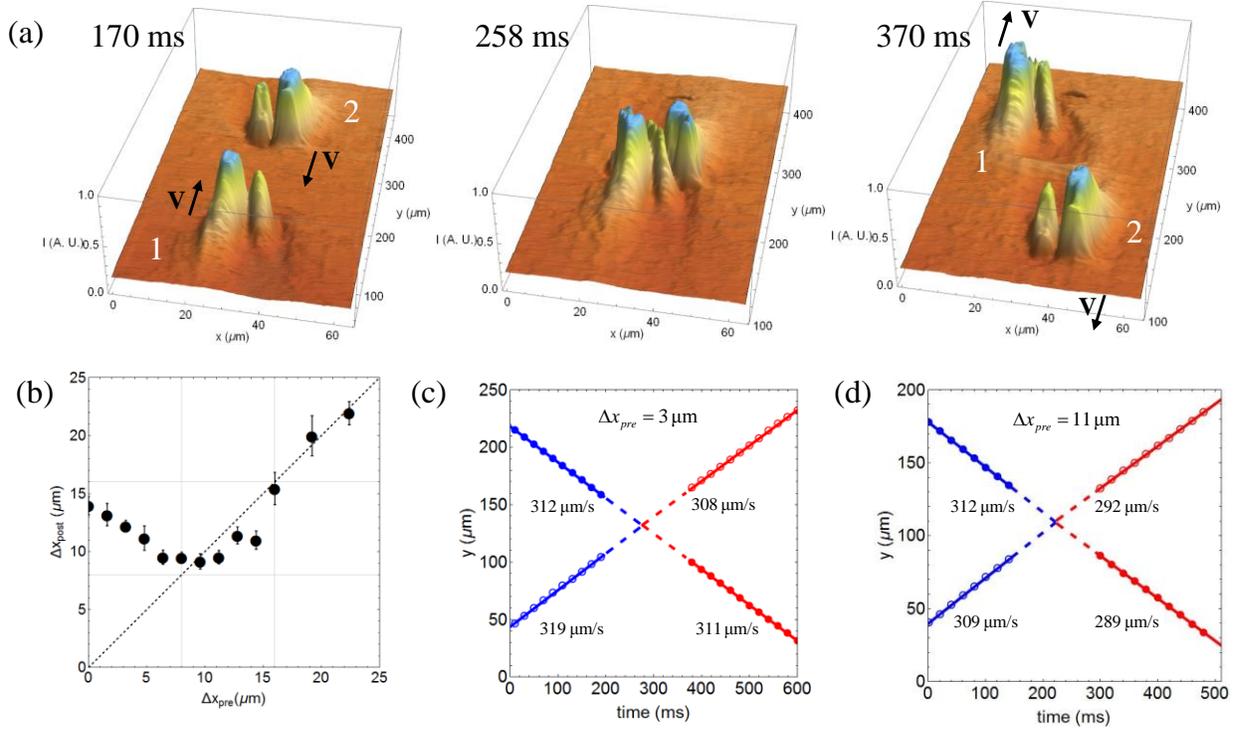

**Fig. 3. Collisions and interactions of solitons.** (a) 3D plots of light intensity of solitons 1 and 2 vs $x, y$ coordinates before, during, and after collision; $\Delta x_{pre} < w/2$; collision results in repulsion, $\Delta x_{post} > \Delta x_{pre}$ ($U = 55.2$ V, $f = 800$ Hz, $T = 50\,°\text{C}$, $d = 8.0\,\mu\text{m}$). (b) Post-collision separation $\Delta x_{post}$ as a function of the impact factor $\Delta x_{pre}$ ($U = 54.8$ V, $f = 400$ Hz, $T = 45\,°\text{C}$, $d = 8.0\,\mu\text{m}$). (c) Time dependence of the $y$-coordinates of two colliding solitons with $\Delta x_{pre} < w/2$; $U = 52.1$ V, $f = 450$ Hz, $T = 45\,°\text{C}$, $d = 7.5\,\mu\text{m}$. (d) The same for solitons with $w/2 < \Delta x_{pre} < w$; the same conditions as in (c). The lines represent linear fitting used to calculate the velocities indicated in the figures.

**Collisions and interactions of solitons.** Solitons moving towards each other can collide, Fig.3. The most frequent scenario is that they reshape during the collision and then recover their structure and constant velocity, which might be different from the pre-collision velocity, Fig.3, Extended Data Fig.6. They also show short-range interactions that depend on the impact parameter $\Delta x_{pre}$. When $\Delta x_{pre} > w$, the solitons pass each other without noticeable perturbations. Solitons that collide "head-to-head", $\Delta x_{pre} < w/2$, Fig.3a, show a repulsion along the $x$-axis, as their post-collision separation increases, $\Delta x_{post} > \Delta x_{pre}$, Fig.3b. Solitons with $w/2 < \Delta x_{pre} < w$ attract each other along the $x$-axis, as $\Delta x_{post} < \Delta x_{pre}$, Fig. 3b. The solitons completely recover their shape after collisions, Fig. 3a and Extended Data Fig.6. Figures 3c,d present trajectories of soliton pairs with $\Delta x_{pre} < w/2$ and $w/2 < \Delta x_{pre} < w$, respectively. The solitons show constant pre- and post-collision velocities.

Other outcomes of pair collisions include annihilation (Extended Data Fig.7) and disappearance of one soliton. Collision with an irregularity in a cell can result in soliton's death (Extended Data Fig.8) or reflection (Extended Data Fig.9). Edges of electrodes can cause both nucleation and disappearance (Extended Data Fig.5).

**Discussion.** The electrically driven dynamic particle-like soliton represents a propagating wave of director deformation that is self-trapped along all three spatial coordinates. The solitons preserve their shape during rapid motion over long distances and after collisions. Their interactions involve short-range attraction and repulsion. The ability to propel along the $y$-axis is caused by two factors: (a) up and down tilts of the director at the head of soliton with the frequency of the applied AC electric field and (b) broken fore-aft symmetry of the director structure in which the sign of the azimuthal angle $\varphi$ does not change when the polar angle $\theta$ oscillates around zero.

The electrically induced solitons represent a novel electrohydrodynamic effect in nematics. The observation is distinct in two aspects. First, although the electroconvection patterns have been observed by many groups[2], there were no reports of self-trapped dynamic soliton waves. The closest structures are "worms" [14], representing elongated formations parallel to the initial director $\hat{\mathbf{n}}_0 = (1,0,0)$ with a finite width but no well-defined length. Another localized structures are islands filled with stripe domains [15]. The islands grow upon the voltage increase till they fill the entire area with the stripe pattern; they do not move laterally. Both worms and islands occur in nematics with $\Delta\varepsilon < 0$ and $\Delta\sigma > 0$ [16]. This combination fits the classic Carr-Helfrich mechanism [1] of electroconvection, according to which the positive anisotropy of conductivity is responsible for destabilization of the planar state. A spatial director fluctuation in presence of an electric field produces space charges and corresponding Coulomb forces which cause instability, usually in the form of space-filling periodic stripes.

The second aspect is that the solitons are observed in the so-called (-,-) nematic with $\Delta\varepsilon < 0$ and $\Delta\sigma < 0$. The Carr-Helfrich mechanism is irrelevant for (-,-) nematics, as dielectric and conductivity torques could only stabilize the planar state. There are two other mechanisms of coupling between the electric field and the director: surface polarization and flexoelectric polarization. Surface polarization can occur at the boundaries of nematic cells because of preferred alignment of molecular dipoles. In presence of the electric field, it can cause surface deformations [17] but not the bulk deformations, which contradicts the observed properties of the solitons with maximum tilts in the middle of the cell. Surface polarization as a reason for solitons is thus ruled out. The main mechanism of soliton formation should be related to flexoelectric polarization.

The flexopolarization $\mathbf{P}_{fl} = e_1\hat{\mathbf{n}}\operatorname{div}\hat{\mathbf{n}} - e_3\hat{\mathbf{n}}\times\operatorname{curl}\hat{\mathbf{n}}$, where $e_1$ and $e_3$ are the flexoelectric coefficients, [6] occurs whenever the director contains splay and bend distortions, which is the case of the solitons, Fig.1b,h. Flexopolarization leads to spatial charge density $\rho_{fl} = \nabla \cdot \mathbf{P}_{fl}$ and a corresponding Coulomb force $\rho_{fl}\mathbf{E}$ [18]. Flexopolarization was suggested as a prime mechanism of the so-called "nonstandard" stripe electroconvection in (-,-) nematics [19], similar to Fig.2c. There is only one instance of localized formations in (-,-) nematics, in the shape of "butterflies" that could move in the plane of the cell, as observed by Brand et al [20]. The report [20], however, did not reveal the director structure of formations nor the direction of motion.

The periodic oscillation of the soliton's director with the frequency $f$, Fig.1g,h, offers a strong support of the idea that the flexoelectric torque $\mathbf{\Gamma}_{fl} = \mathbf{P}_{fl} \times \mathbf{E}$ is responsible for solitons formation, since it is linear in the field, $\mathbf{\Gamma}_{fl} = \mathbf{P}_{fl} \times \mathbf{E}$. Further support is provided by the similar symmetry properties of flexoelectric stripes described by Krekhov et al [18] and our solitons. Namely, the reversal of the electric field polarity causes reversal of the polar angle $\theta$ but leaves intact the sign of $\varphi$. The theory [18] predicts the same type of frequency dependence of $U_{stripes}$ as in Fig.2a. The flexoelectric force driving a soliton can be estimated roughly as $e^*U \sim 5 \times 10^{-10}$ N, where $e^* \sim 10^{-11}$ C/m (Ref [1]) is an effective flexocoefficient and $U = 50$ V. This estimate agrees very well with the expected viscous drag force, $\sim R\eta v \sim 5 \times 10^{-10}$ N, where $R \sim 10$ μm is the effective radius of the soliton, $\eta \approx 60$ mPa·s is the viscosity of CCN-47 and $v \approx 0.8 \times 10^{-3}$ m/s, that corresponds to $U = 50$ V, Fig. 2b.

The solitons are not unique to CCN-47. We observe them also in other materials, such as ZLI-2806 (Merck). M.H. Godinho and C. Rosenblatt observed similar formations in their nematic systems (unpublished results, private communications). A critical condition for the soliton formation is relatively low conductivity. In the studied CCN-47, $\sigma_{\parallel}, \sigma_{\perp} \approx (5-6) \times 10^{-9}$ $\Omega^{-1}$m$^{-1}$, which is much smaller than the conductivity $10^{-7}\,\Omega^{-1}\mathrm{m}^{-1}$ usually reported in the studies of stripe electroconvective patterns in (-,-) nematics [21]. When the conductivity of CCN-47 is increased to $10^{-7}\,\Omega^{-1}\mathrm{m}^{-1}$ and the concentration of ions is raised above $10^{21}\,\mathrm{m}^{-3}$ by addition of salts such as tetrabuthylammonium bromide, the solitons are suppressed and the nematic undergoes a direct transition to stripe domains. The condition of low conductivity might explain why the solitons have not been reported before.

To summarize, we demonstrated that the electric field can cause a novel structural response of a nematic liquid crystal representing propagating solitary waves that are self-trapped along all three spatial dimensions. The dynamics is caused by periodic oscillations of the director with the same frequency as the frequency of the electric field which suggests that the underlying mechanism is related to flexoelectric polarization. The presented rich dynamic behavior of these 3D particle-like solitons in a system that is relatively easy to control should open the door to a broad range of further studies. Of especial importance would be the study of director deformations during pairwise collisions, dependency of the soliton properties on the control parameters and collective behavior of solitons at concentrations that are sufficiently high for multiparticle interactions. The observed solitons pose a formidable problem for a detailed theoretical description that should take into account a complex mix of dielectric, conductive, flexoelectric, elastic, and anisotropic viscous forces.

**Methods**

**Materials.** The studied nematic is a one-component 4'-butyl-4-heptyl-bicyclohexyl-4-carbonitrile CCN-47 (Nematel GmbH). The phase diagram of CCN-47 is smectic A 29.9 °C nematic 58.5 °C isotropic phase. We confirmed the conclusion of Dhara and Madhusudana [13] that the anisotropy of both the permittivity and conductivity of CCN-47 is negative, by performing independent measurements using an LCR meter 4284A (Hewlett-Packard) and cells with planar (alignment agent polyimide PI-2555, HD MicroSystems) and homeotropic (polyimide SE1211) alignment. In particular, at $45\,°C$ and 4kHz, $\varepsilon_\perp \approx 8.8$, $\varepsilon_\parallel \approx 4.6$, $\sigma_\perp \approx 6.1 \times 10^{-9}\ \Omega^{-1}\mathrm{m}^{-1}$ and $\sigma_\parallel \approx 4.9 \times 10^{-9}\ \Omega^{-1}\mathrm{m}^{-1}$; concentration of ions is $4 \times 10^{20}\ \mathrm{m}^{-3}$. The effective viscosities of CCN-47 are $\eta_\parallel = 57\ \mathrm{mPa\cdot s}$ and $\eta_\perp = 58\ \mathrm{mPa\cdot s}$ for a motion parallel and perpendicular to the director, respectively, as determined by tracking Brownian motion of polystyrene colloidal spheres of diameter 5 μm and tangential surface anchoring. [22]

**Generation of solitons.** The cells were aligned in a planar fashion by using rubbed layers of PI-2555. The directions of rubbing at two opposite plates are antiparallel. The glass plates contain transparent indium tin oxide (ITO) electrodes of area $5 \times 5\ \mathrm{mm}^2$. The temperature of the cell is controlled with a Linkam LTS350 hot stage and a Linkam TMS94 controller. The AC electric field was applied using a waveform generator (Stanford Research Systems, Model DS345) and amplifier (Krohn-hite Corporation, Model 7602).

**Optical characterization of solitons.** The material is of a low birefringence, $\Delta n = n_e - n_o \approx 0.03$, Ref. [13], ($n_e$ and $n_o$ are the extraordinary and ordinary refraction indices, respectively) which makes it convenient to explore the solitons and the associated director deformations using standard polarizing microscopy with a wave-plate (red plate) optical compensator and fluorescence confocal polarizing microscopy [23]. The director distortions are characterized by analyzing the transmitted light intensity $I = I_0 \sin^2(2\varphi - 2\psi)\sin^2(\Gamma/2)$ determined by optical phase retardance

$$\Gamma = \frac{2\pi \Delta n d}{\lambda}\left(1 - \sin^2\theta\cos^2\varphi - \frac{\sin\alpha\sin 2\theta\cos\varphi}{\bar{n}} - \frac{\sin^2\alpha\left(\sin^2\theta\cos^2\varphi - \cos 2\theta\right)}{\bar{n}^2}\right)$$ and by the

angle $\psi$ between y-axis and the polarizer; here $\bar{n} = (n_e + n_o)/2$ is the average refractive index. When the director is along the polarizer or the analyzer, the texture is dark. The azimuthal angle $\varphi$ is determined by rotating the cell in the normal incidence geometry $(\alpha = 0)$ around the $z$-axis (scaning $\psi$). Knowing $\varphi$, we calculate the maximum polar tilt $\theta_{max}$ from the minimum light transmittance.

We used polarizing Nikon TE2000 inverted microscope equipped with two cameras: Emergent HR20000 with resolution $5120 \times 3840$ pixels and the frame rate up to 1000 frames/s and MotionBLITZ EOSens mini1 (Mikrotron GmbH) with the frame rate up to 6000 frames/s. The coordinate $y$ of the solitons in Fig.3c,d is defined as the location of the maximum light intensity transmitted through the soliton observed between the two polarizers and an optical compensator.

To verify that the maximum director distortions in the solitons are located in the middle of the cell, we used fluorescence confocal polarizing microscope (FCPM) based on Olympus Fluoview BX-50. [24] CCN-47 is doped with a small amount (0.1 wt%) of a fluorescent dye n,n'-bis(2,5-di-tert-butylphenyl)-3,4,9,10-perylenedicarboximide (BTBP), purchased from Molecular Probes, the molecules of which are elongated and align parallel to the director. FCPM uses a linearly polarized laser beam to probe the specimen under normal incidence. The fluorescent signal is maximum when the light polarizations is parallel to the transition dipole of BTBP [24] and minimum when the two are orthogonal. The scanning focused laser beam is linearly polarized along the $y$-axis. Generation of solitons produced an increase of the fluorescent signal from the middle of the cell which confirm director distortions in this region.

Supplementary materials are available in the online version of the paper.

Reprints and permissions information is available at www.nature.com/reprints

**Acknowledgements**

We are thankful to A. Buka, M. C. Calderer, N.A. Clark, M.A. Glaser, M.H. Godinho, D. Golovaty, and C. Rosenblatt for useful discussions. This work is supported by NSF grants DMS-1729509 and DMR-1507637.

**Author Contributions:**

BL, VB, RX, and SP performed the experimental studies. TT measured the viscosity of CCN-47. SVS derived the expression for light transmission, BL, VB, RX, SP, SVS, and ODL analyzed the data and discussed the mechanisms. ODL directed the research and wrote the manuscript with an input from all coauthors.

**Author Information:**

Correspondence and requests for materials should be addressed to olavrent@kent.edu.

# Extended Data

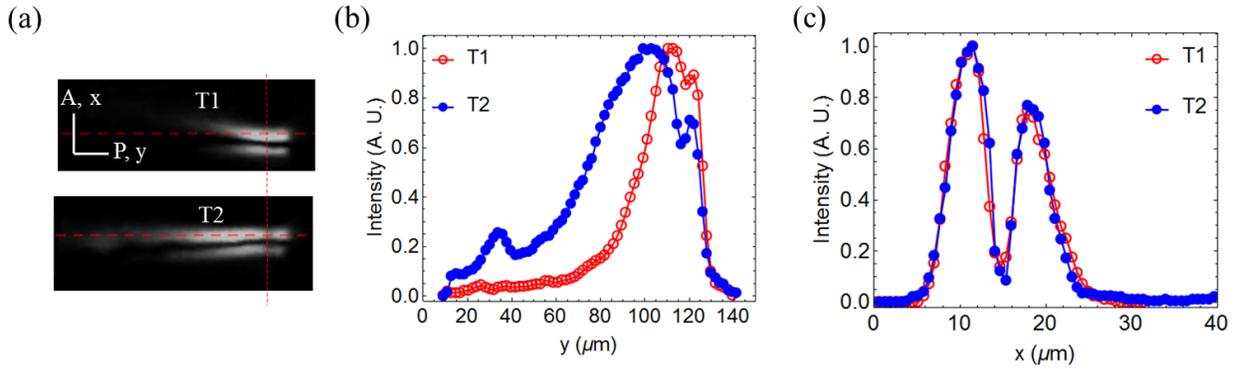

**Extended Data Fig. 1. Voltage dependences of the length and width of solitons.** (a) Polarizing microscopy textures of two solitons, T1 ($f = 500$ Hz, $U = 41.2$ V) and T2 (850 Hz, 56.4 V) in a cell with $d = 7.8$ µm; t; (b) profiles of T1 and T2 solitons along the $y$-axis plotted as light intensity vs. $y$-coordinate along the horizontal lines shown in part (a). The light intensity was normalized to the same maximum for a better comparison; (c) profile of T1 and T2 solitons along the $x$-axis plotted as light intensity vs. $x$-coordinate along the vertical lines shown in part (a); the width shows practically no dependence on the applied voltage.

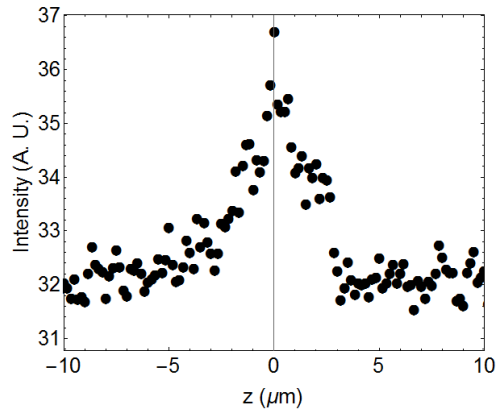

**Extended Data Fig. 2. Fluorescent signal measured by FCPM across the cell**. CCN-47 cell under an AC field, $U = 42.0$ V, $f = 200$ Hz. The background fluorescence signal from a cell at zero voltage is subtracted from the plot. The cell thickness $d = 8.0$ µm, temperature $T = 35\,°C$. The middle of the cell is set at $z = 0$.

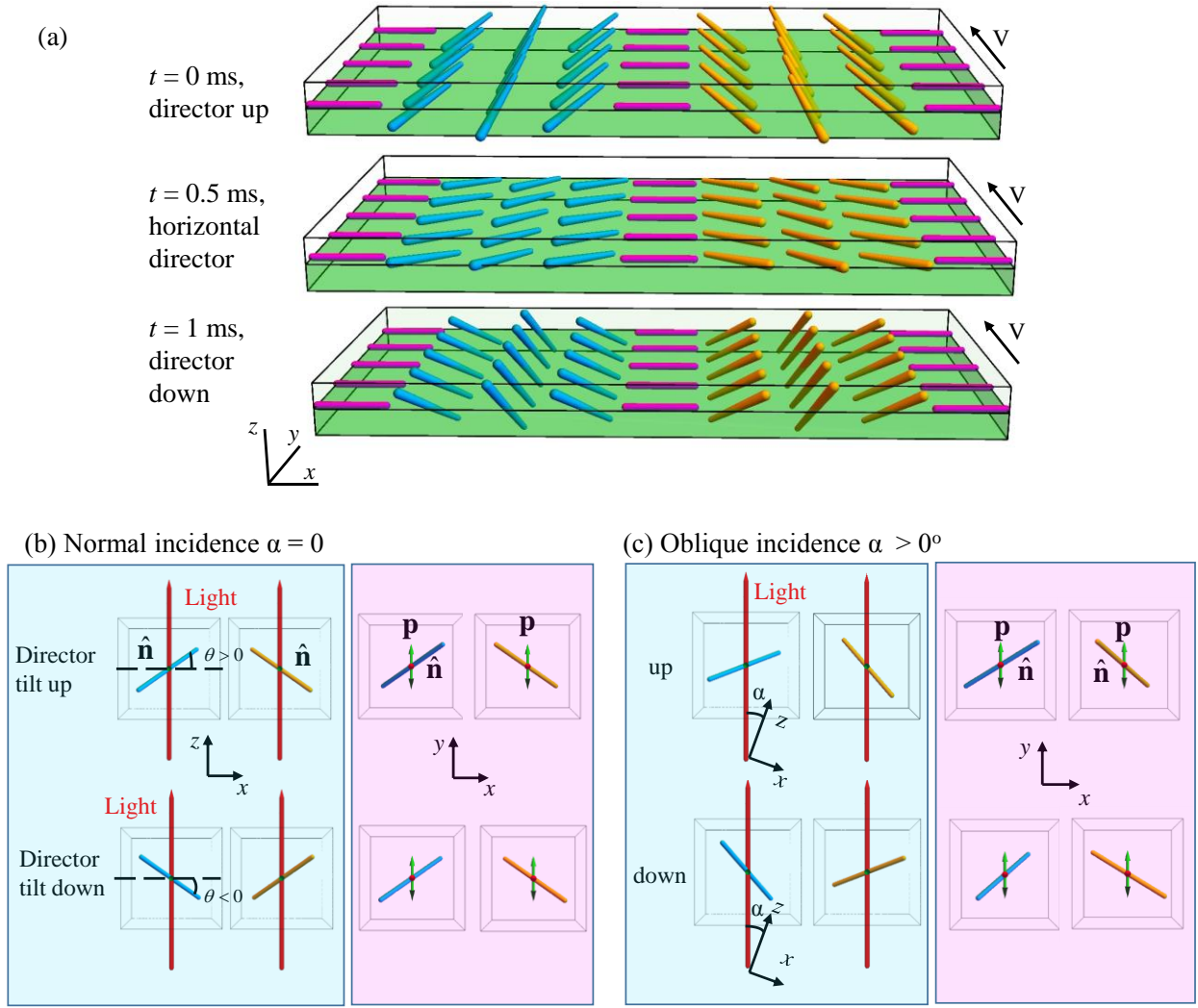

**Extended Data Fig. 3. Polar tilts of the director within the head of the soliton.** (a) Scheme of the director; (b) Normal incidence of the light beam polarized along the $y$-axis; left and right parts of the soliton produce the same angle between the light polarization **p** and the local director; (c) Oblique incidence of the polarized beam results in a different light transmittance at a given moment of time through the left and right sides of the soliton because the angle between the local director and **p** is different in these two sides.

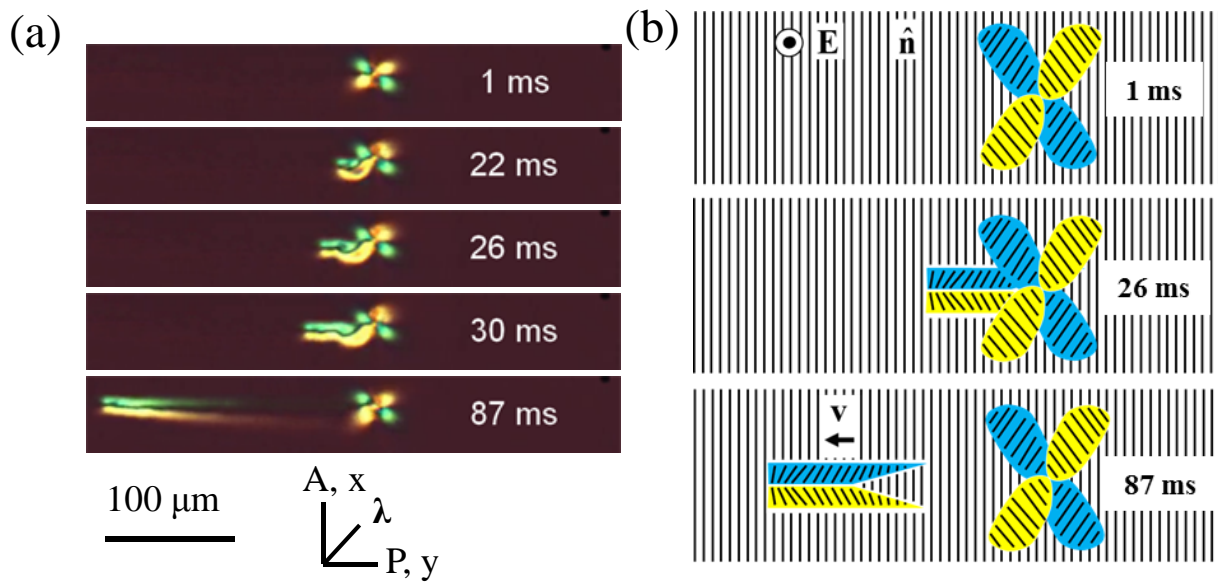

**Extended Data Fig. 4. Nucleation of a soliton at an irregularity** (a) $U = 87.4$ V, $f = 1000$ Hz, $T = 45\,°C$, $d = 8.2$ μm; (b) director scheme of nucleation;

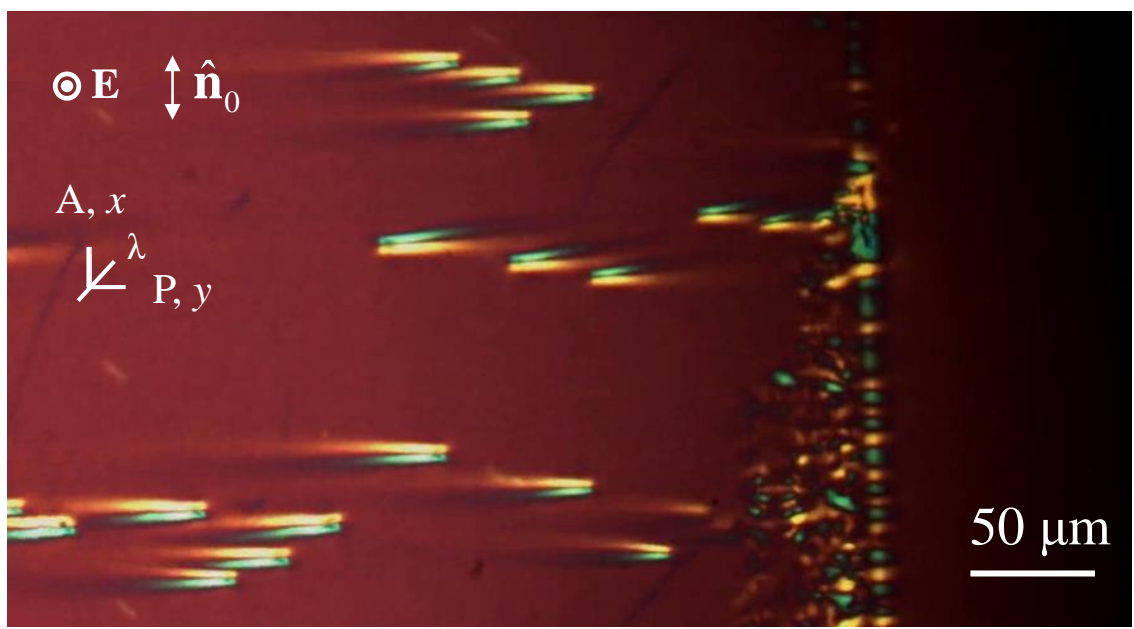

**Extended Data Fig. 5. Nucleation and disappearance of solitons at the edge of the electrode area.** $U = 72.2$ V, $f = 1000$ Hz, $T = 50\,°C$, $d = 8.0$ μm.

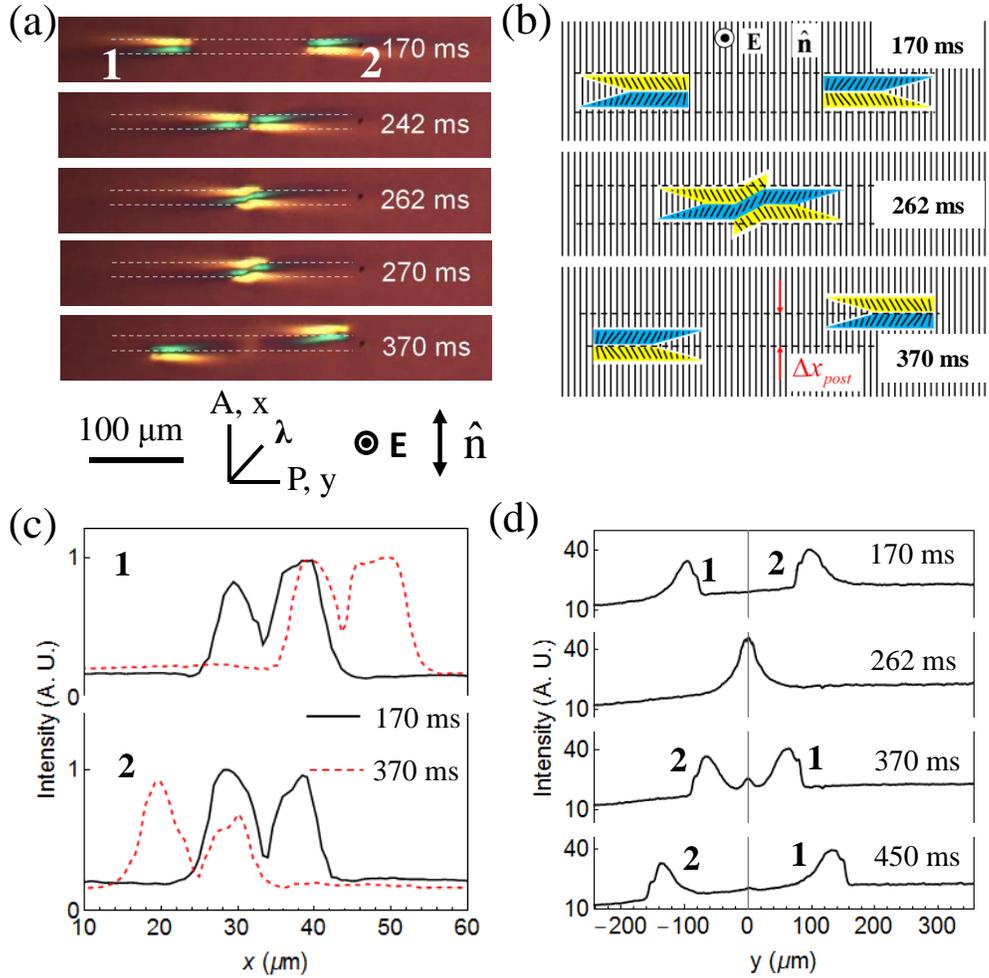

**Extended Data Fig. 6. Collisions and interactions of solitons.** (a) Polarizing microscopy textures of collision of solitons 1 and 2, with $\Delta x_{pre} < w/2$; $U = 55.2$ V, $f = 800$ Hz, $T = 50$ °C, $d = 8.0$ μm; (b) corresponding schemes; (c) Transmitted light intensity for solitons 1 and 2 as a function of $x$ coordinate before and after collision (measured along the line i=15 in Fig.1b). The left and right parts of the soliton are of a different light intensity because of the presence of the red plate that imparts the blue and yellow colors on the two sides of the tuxedos. (d) Profiles of solitons 1 and 2 as a function of $y$ coordinate before and after collision, obtained by integrating light intensity transmitted through the soliton width. Both $x$- and $y$-profiles of the solitons show a remarkable recovery after the collision.

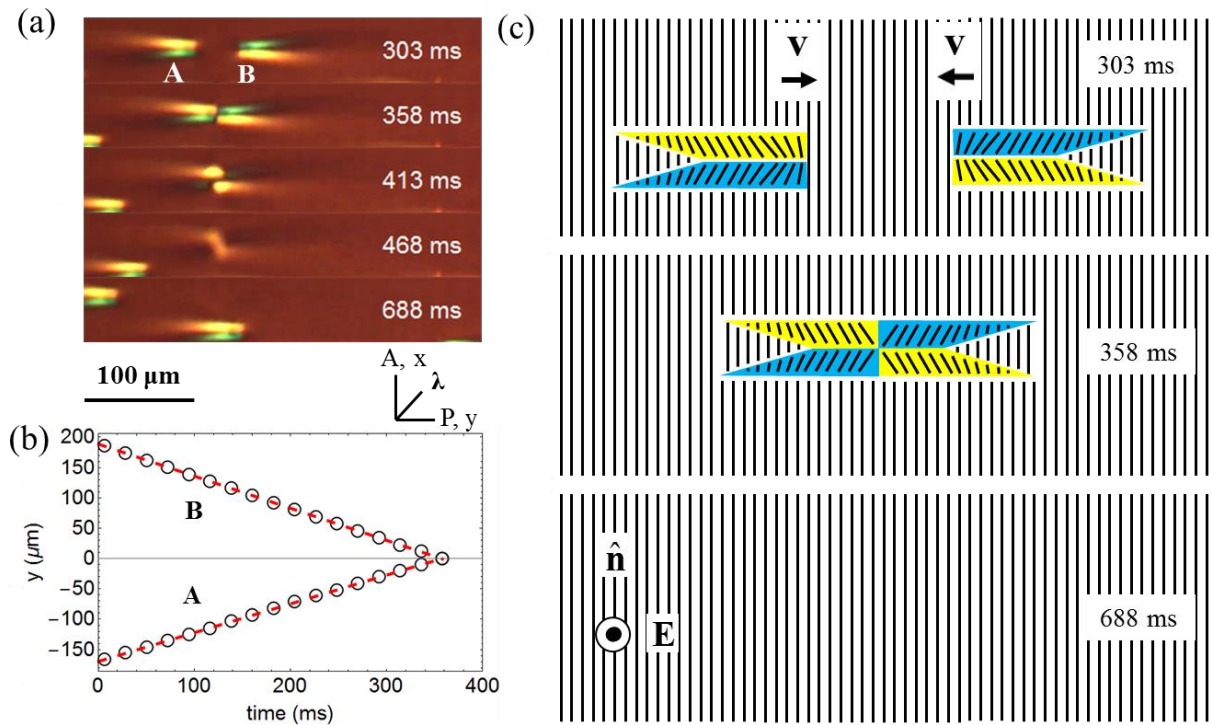

**Extended Data Fig. 7. Annihilation of solitons in pairwise interaction.** (a) Polarizing microscopy of time sequence of annihilation of two solitons; $U = 45.1\,\text{V}$, $f = 600\,\text{Hz}$, $T = 50\,°\text{C}$, $d = 8.0\,\mu\text{m}$; (b) location of soliton (determined as the coordinate of the peak of light transmittance) vs. time.

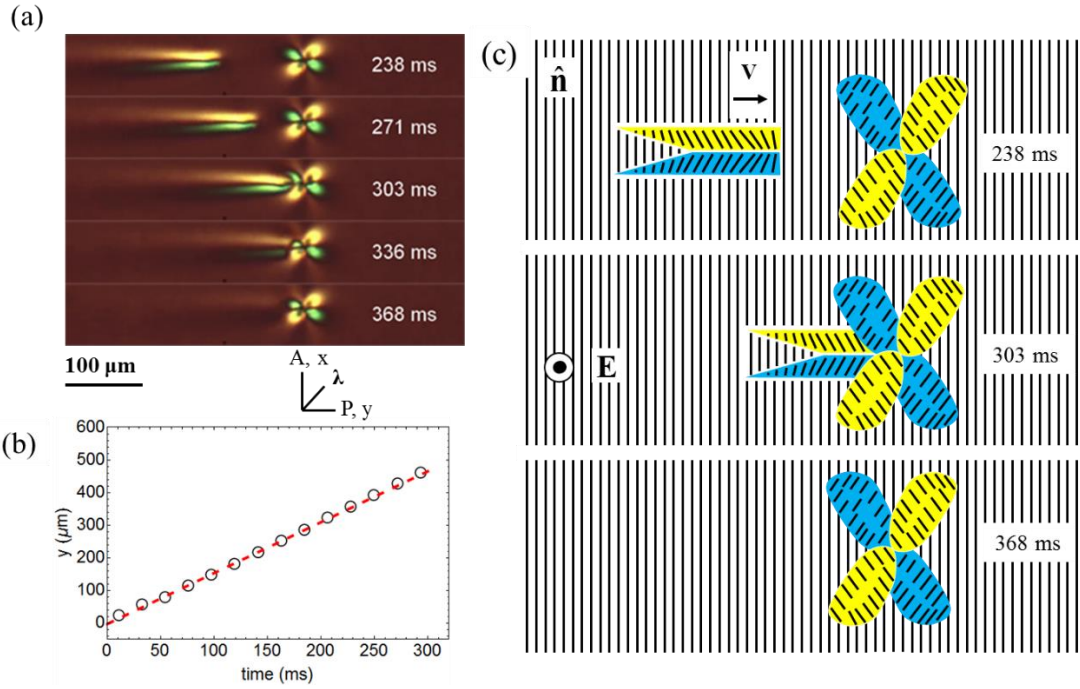

**Extended Data Fig. 8. Disappearance of a soliton at an irregularity in CCN-47 cell.** $U = 65.6$ V, $f = 800$ Hz, $T = 50\,°C$, $d = 7.7\,\mu m$.

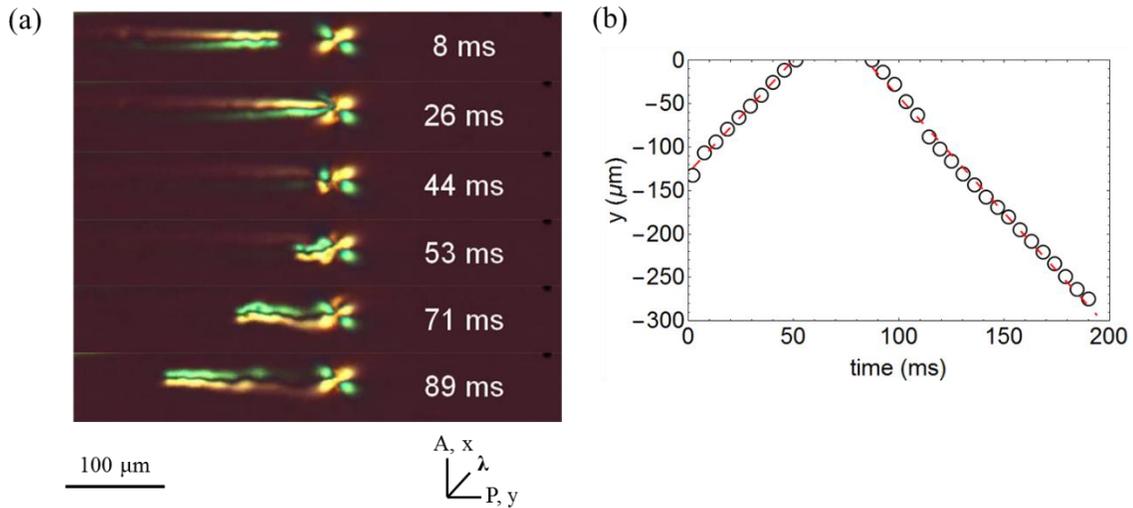

**Extended Data Fig. 9. Reflection of a soliton at an irregularity in a CCN-47 cell.** $U = 87.4$ V, $f = 1000$ Hz, $T = 45\,°C$, $d = 8.2\,\mu m$.

22